\documentclass[twocolumn]{aastex631}
\usepackage{amsmath}
\usepackage{subfigure}
\usepackage{bm}

\usepackage{graphicx}
\usepackage{multirow}
\usepackage{ulem}

\begin{document}

\title{Spatial distribution of secondary electrons' Synchrotron emission: property and implication}

\author[0000-0003-2245-4364]{Qi-Zuo Wu}
\affiliation{School of Astronomy and Space Science, Nanjing University, 163 Xianlin Avenue,\\
Nanjing 210023, People’s Republic of China}
\affiliation{Key Laboratory of Modern Astronomy and Astrophysics, \\Nanjing University, Ministry of Education,
Nanjing 210023, People’s Republic of China}

\author[0000-0003-1576-0961]{Ruo-Yu Liu}
\affiliation{School of Astronomy and Space Science, Nanjing University, 163 Xianlin Avenue,\\
Nanjing 210023, People’s Republic of China}
\affiliation{Key Laboratory of Modern Astronomy and Astrophysics, \\Nanjing University, Ministry of Education,
Nanjing 210023, People’s Republic of China}
\affiliation{Tianfu Cosmic Ray Research Center,
 Chengdu 610000, Sichuan, People’s Republic of China}

\correspondingauthor{Ruo-Yu Liu}
\email{ryliu@nju.edu.cn}


\begin{abstract}
Galactic $\gamma$-ray sources can be produced by either high-energy protons via proton-proton collisions or electrons/positrons via inverse Compton scattering. Distinguishing between the hadronic and leptonic origin of $\gamma$-ray emission in Galactic sources remains challenging. Measurements of non-thermal X-ray spectra of these sources, which could originate from primary electrons in the leptonic scenario or secondary electrons/positrons in the hadronic scenario, have been suggested as an efficient way of discriminating between these scenarios. In this work, we investigate the morphology of the X-ray emission from secondary electrons/positrons. By calculating the surface brightness profile and the photon index profile of X-ray emission, we find that secondary electrons produce a distinctively flat X-ray surface brightness profile. Our results suggest that, in addition to the X-ray spectrum, the X-ray morphology is crucial to determine the radiation mechanism of ultrahigh-energy $\gamma$-ray sources and help to identify sources of PeV cosmic rays.
\end{abstract}



\section{Introduction} \label{sec:intro}

The origin of PeV cosmic rays (CR) is a long-standing puzzle in high-energy astrophysics.
Ultrahigh-energy $\gamma$-ray (UHE, $E>100\,$TeV) emission is an effective probe of PeV CR accelerators. So far, there are more than 40 UHE $\gamma$-ray sources mainly detected by the Large High Altitude Air Shower Observatory (LHAASO),  and by the High Altitude Water Cherenkov observatory (HAWC). The astrophysical counterparts of these UHE $\gamma$-ray sources include supernova remnants (SNR), pulsar wind nebulae (PWN), young massive clusters (YMCs), and microquasars, and all these source population have been suggested to be potential capable of accelerating protons to PeV or higher energies. On the other hand, UHE $\gamma$-ray emission can arise either from proton-proton ($pp$) collisions or from the inverse Compton scattering (ICS) of electrons interacting with the interstellar radiation field (ISRF) and the cosmic microwave background (CMB), categorizing UHE $\gamma$-ray emissions into hadronic and leptonic origins. Understanding the radiation mechanism of these sources is crucial to pinpoint sources of PeV CRs\footnote{Note that, UHE $\gamma$-ray sources which are dominated by the leptonic radiation may still be PeV proton accelerators. Hadronic radiation may be subdominant because of a lack of dense matter target inside or around accelerators.}.

It is challenging to differentiate the radiation mechanism based solely on $\gamma$-ray observations \citep{2024A&A...691A..89A,2021ApJ...912..158M,2024ApJ...970L..21A}. From the perspective of the spectrum, the Klein-Nishina (KN) effect can soften the $\gamma$-ray spectrum of ICS in the UHE band, so observation of a relatively hard power-law spectrum up to an energy above 100\,TeV is usually considered as an indication of the hadronic origin. However, if the electron spectrum is hard, for instance, $dN/dE_e \propto E^{-1}$, and an appropriate cutoff energy in the high-energy end of the spectrum is given, the resulting ICS spectrum may approximately form a power-law distribution of an index of $-2$ in the UHE band \citep{Liu2021}. Moreover, according to the spectra of UHE $\gamma$-ray sources in the first LHAASO catalog, the photon indexes of most of these UHE sources are steeper than $-3$ above 25\,TeV, and both leptonic and hadronic scenarios may work (e.g., fitting to the spectrum of LHAASO~J1908+0621 shown in \citealt{2021Natur.594...33C}). Measurements of $\gamma$-ray morphology may provide additional hints to the radiation mechanism, because $\gamma$-ray sources of the leptonic origin are expected to show a different surface brightness profile and a different energy-dependent behavior from those of the hadronic origin due to cooling of electrons. However, given the possible contamination from nearby sources and due to the limited angular resolution of $\sim 0.2^\circ$ of the instruments, accurate measurements are not always available. 
The situation will become even more complex if the source itself contains two or more particle acceleration zones and hence two or more particle components.

Therefore, multi-wavelength observations are crucial to identify PeV CR sources. In particular, X-ray observation can offer critical information on the $\gamma$-ray radiation mechanism. Electrons emitting UHE $\gamma$ rays via ICS may generate X-ray emission via the synchrotron radiation. If the nonthermal X-ray flux and the $\gamma$-ray flux of a source are both detected, we can test whether the leptonic scenario can work with a reasonable magnetic field, although the latter is often a quantity of large uncertainty. On the other hand, non-thermal X-ray emission is also expected in the hadronic scenario, arising from the synchrotron radiation of secondary electrons/positrons (hereafter we do not distinguish positrons from electrons for simplicity). \citet{2020ApJ...903...61C} found that the energy distributions of synchrotron radiation from secondary electrons are shallower than (or flatter/harder than) parent electrons and $\gamma$-rays in the cutoff region, which could potentially serve as a promising diagnostic approach for identifying the hadronic origin of the $\gamma$-ray radiation. However, for PeVatrons, the synchrotron spectral cutoff would generally appear in the hard X-ray band or soft $\gamma$-ray band, which is not easy to detect with the capabilities of current-generation hard X-ray/soft $\gamma$-ray instruments.

X-ray instruments, which generally have good angular resolution better than 1 arcmin, may offer accurate measurements of the source morphology. In addition to the X-ray spectrum, the morphological information of the source in the X-ray band, such as the surface brightness profile (SBP) and the photon index profile, also carries crucial information about particle distribution and transport. In our previous study \citep{Wuqz2024}, we modeled the $\gamma$-ray SBPs and spectra measured in three pulsar halos under different models. We then calculate corresponding X-ray SBPs and the photon index profiles under the models and found that future X-ray observations may distinguish different models of pulsar halos. Since X-ray emission is also expected in the hadronic scenario, we aim to study quantitatively whether the X-ray SBP and the photon index profile in the hadronic scenario present distinct features from the leptonic scenario, and how we can use these measurements as a complementary diagnostic tool to distinguish the radiation mechanism of UHE $\gamma$ rays.

The rest of the paper is structured as follows: In Section \ref{sec:method}, we introduce the propagation and multi-wavelength radiation of CRs for hadronic and leptonic scenarios. Focusing on the synchrotron radiation of secondary electrons, we analyze our calculations and discuss their application in distinguishing $\gamma$-ray and X-ray origins in Section \ref{sec:result}. Results are discussed and concluded in Section \ref{sec:dis}.

\section{Particle transport and radiation} \label{sec:method}
We assume the protons/electrons are injected from a point-like source and isotropically diffuse in the ambient medium with uniform density $n_{0}=10~ \text{cm}^{-3}$ , which is close to the value of cold interstellar medium (ISM). We assume a spatially independent diffusion coefficient with Kraichnan-type: $D(E)=D_0 ({E}/{1~\text{GeV}})^{1/2}$ for both hadronic and leptonic scenarios. The source is assumed to be located at $d=5\,$kpc away from Earth, and the age of the system is assumed to be $T_{\text{age}}=2\,$Myr, following the age of YMCs which have been suggested to be PeVatron or super PeVatron \citep{2021NatAs...5..465A, 2024SciBu..69..449L}.

\subsection{Hadronic scenario}
For the hadronic scenario, we assume a constant injection of protons with a power-law spectrum:
\begin{equation}
\resizebox{0.65\columnwidth}{!}{$
Q_p(E_p) = N_p^{\text{inj}} \times
    \left(\frac{E_p}{100\,\text{TeV}}\right)^{-s_{p}}
$}
\end{equation}
where $N_p^{\text{inj}}$ can be obtained according to the proton injection luminosity $L_{p}(E_p\textgreater 100 ~ \rm TeV)$ by $\int E_pQ_p(E_p)dE_p$, $s_p$ is the power-law index of protons. The average energy loss time of protons due to the $pp$ collisions is almost energy-independent $\tau_{\text{pp}} \sim 10^{7} (n_{0}/ \text{cm}^{-3})^{-1}$\,yr, which is much longer than the diffusion timescale of high-energy protons (which are relevant with this study) within the source. Hence, we neglect the energy loss of protons. Setting $t=0$ as the beginning of particle injection, the distribution of protons at time $t$ is given by \citep{1995PhRvD..52.3265A}:
\begin{equation}
n_p(E_p, r, t) = \frac{Q_p}{4 \pi D r} \cdot \mathrm{erfc}\left(\frac{r}{2 \sqrt{D t}}\right)
\label{eq:np}
\end{equation}
where {\it erfc} is the error function and equals to 1 when $r \gg 2 \sqrt{D t}$, thus exhibits a $1/r$ profile for the proton density. As protons diffuse outward and collide with the surrounding medium, a variety of secondary particles are produced, and here we focus on the $\gamma$-ray and secondary electrons and their synchrotron radiation.

The total inelastic cross section of $pp$ collisions is adopted following the parameterization in \citet{2014PhRvD..90l3014K}. Since we have known the proton density at present $n_{p}(E_{p},r,T_{\text{age}})$, we can directly derive the $\gamma$-ray flux following the formula in \citet{2014PhRvD..90l3014K}. In this work, we use the 'Pythia 8' model. 

For secondary electrons, however, the situation is more complex. As protons diffuse, secondary electrons are continuously produced at every location where protons interact with gas. The produced electrons will also diffuse and suffer energy loss. To obtain the final electron distribution, we first calculate the injection rate of secondary electrons using the parameterizations of energy spectra given by \citet{2006PhRvD..74c4018K}:
\begin{equation}
\resizebox{\columnwidth}{!}{$
\begin{aligned}
\frac{dN_e}{dE_edVdt}&=Q_{e}(E_e, r, t) =   c \int_{100 \text{GeV}}^{E_p^{\text{max}}} n_{\text{gas}}(r) \sigma_{\text{inel}}(E_p) \\
&  \times n_{p}(E_p, r, t)  F_e(E_e/E_p, E_p) \frac{dE_p}{E_p}
\end{aligned}
$}
\end{equation}
where $c$ is the speed of light, $E_{p}^{\text{max}}$ is set to 100~PeV, $n_{\text{gas}}=n_{0}$ is the gas density, and we assume the gas is distributed within a radius of 200~pc. $n_{p}(E_p, r, t)$ is the proton density derived from Equation (\ref{eq:np}). $F_e(E_e/E_p, E_p)$ is the kernel function in \citet{2006PhRvD..74c4018K}. These secondary electrons will not only propagate outward in space, but also lose energy through synchrotron radiation and ICS (including the KN effect), at an energy-loss rate which can be approximately described by \citet{2021Sci...373..425L}:
\begin{equation}
\resizebox{\columnwidth}{!}{$
\dot{E}_{\mathrm{e}} = -\frac{4}{3}\sigma_{\mathrm{T}}c\left( \frac{E_{\mathrm{e}}}{m_{\mathrm{e}}c^2} \right)^2\left\{ U_{\mathrm{B}}+\sum_i U_{\mathrm{ph},i}\,/\left[ 1+\left( \frac{2.82kT_{\mathrm{i}}E_{\mathrm{e}}}{m_{\mathrm{e}}^2c^4} \right)^{0.6} \right]^\frac{1.9}{0.6} \right\}
$}
\end{equation}
where $\sigma_{\rm T}$ is the Thomson cross-section, $m_{\rm e}$ is the elecctron mass, $U_{\rm B}=B^2/8\pi$ is the energy density of magnetic field and $U_{{\rm ph},i}$ is the $i$th component of energy density of ISRF. In this work, we adopt the cosmic microwave background (CMB), an infrared radiation field ($T = 30\,$K and $U = 4.8\times10^{-13}\, \rm erg~cm^{-3}$) and a star light radiation field ($T = 5000 \rm K$ and $U =  4.8\times10^{-13}$ $\rm erg$ $\rm cm^{-3}$). For the hadronic scenario, we consider a stronger magnetic field (a few tens $\mu$G to mG), which is consistent with the value near the Galactic center (See \citealt{2009A&A...505.1183F} for a review).  Due to this high B value, and combined with the  Klein-Nishina effect of ICS, the cooling of TeV electrons is primarily dominated by the synchrotron radiation.

Then we can derive the electron density by solving the spherically symmetric diffusion equation :
\begin{equation}
\frac{\partial{n_e}}{\partial{t}}=\frac{1}{r^2}\frac{\partial }{\partial r}\left(r^2 D \frac{\partial n_e}{\partial r}\right)- \frac{\partial}{\partial{E_{\rm e}}}(\dot{E_{\rm e}}n_e)+Q_e(E_e,r,t) 
\label{propagation}
\end{equation}

We solve this equation numerically using the operator splitting technique and semi-implicit method. Then we calculate the corresponding synchrotron radiation following the approximate formulae of the Bessel function given by \citet{2013RAA....13..680F} to improve computational efficiency.

\subsection{Leptonic scenario}
For the leptonic scenario, we adopt a constant continuous injection with an exponential power-law spectrum:
\begin{equation}
\resizebox{0.7\columnwidth}{!}{$
Q_e(E_e) = N_e^{\text{inj}} 
    \left(\frac{E_e}{100\,\text{TeV}}\right)^{-s_{e}} e^{-E_{\mathrm{e}}/E_{\text{cut}}}
$}
\end{equation}
where $N_e^{\text{inj}}$ can be obtained according to the electron injection luminosity $L_{e}(E_e\textgreater 100 ~ \rm TeV)$ by $\int E_eQ_e(E_e)dE_e$, $s_e$ is the power-law index of electrons, and $E_{\text{cut}}$ is the cut-off energy. We assume a point-like injection, and the electron density can be derived from Equation (\ref{propagation}). After obtaining the electron density at present, we calculate their ICS using the simplified formula given by \citet{2014ApJ...783..100K} and also compute their synchrotron radiation. Finally, we get both the SBP and energy spectrum for analysis.

\subsection{Comparison between hadronic and leptonic scenarios}
Before presenting our calculation results, we provide a general comparison between the hadronic and leptonic scenarios, focusing primarily on the relative magnitudes of the X-ray flux.

To begin with, we consider whether the synchrotron radiation from secondary electrons is strong enough to be observed. In $pp$ collisions, $\pi^{\pm}$ and $\pi^0$ mesons are produced in approximately equal proportions, each accounting for about 1/3 of the total energy lost by the relativistic proton \citep{2006PhRvD..74c4018K}. While $\gamma$ rays inherit all of the $\pi^0$ energy, secondary electrons from $\pi^{\pm}$ decays receive about 1/6 of the total energy \citep{2004vhec.book.....A}. Therefore, we expect the flux ratio between the synchrotron radiation of secondary electrons and pionic $\gamma$-ray emission to reach approximately 1:2 at most, in the synchrotron saturation regime (the synchrotron loss timescale is much shorter than other energy loss timescales and the age of the system). Enhancing the magnetic field strength further would only shift the peak energy of the synchrotron spectrum to higher energies \citep{2020ApJ...903...61C}. In contrast, the ratio of synchrotron flux to IC flux in the leptonic scenario can vary significantly depending on the magnetic field strength. Therefore, if the observed X-ray flux is higher than the $\gamma$-ray flux, the X-ray emission cannot primarily come from secondary electrons (even if the $\gamma$-ray is of hadronic origin).

In the following section, the hadronic scenario is associated with a relatively strong magnetic field ($\textgreater 10 \mu \rm G$), while the leptonic scenario is characterized by a relatively weaker magnetic field ($\sim 1 \rm \mu G$). As such, the X-ray flux from both scenarios can be of comparable magnitude.
More specifically, in the hadronic scenario, the secondary electrons and pionic $\gamma$-ray photons from $pp$ collisions are of similar energy, i.e., $E_e \approx 0.5E_{\gamma} \approx 0.05 E_p$. The energy of typical synchrotron photons produced by electrons is given by: 
\begin{equation}
E_{sy} \simeq 2 \left({\frac{E_e}{100 \rm TeV}}\right)^2 \left(\frac{B}{10 \mu \rm G}\right) \rm keV
\end{equation}
For the ICS, the approximate energy relation is given by \citet{2004vhec.book.....A}:
\begin{equation}
E_{e} \simeq 17 E_{\rm ICS, \rm TeV}^{ 0.54+0.046 {\rm log}_{10}(E_{\rm ICS, \rm TeV})} ~ \rm TeV
\end{equation}
For example, 100 TeV electrons correspond to a gamma-ray energy of $\sim$20 TeV. Therefore, we focus on injection of energy around 100\,TeV. 

\section{Result} \label{sec:result}

In this section, we present the properties of secondary electrons and their synchrotron radiation, focusing on the differences in spatial distribution between hadronic and leptonic scenarios. To begin with, we study the 1D radial density and power-law index profile of secondary electrons in the hadronic scenario, and compare them to those of primary electrons in the leptonic scenario. Subsequently, we calculate and present the spatial morphology of the synchrotron radiation produced by these secondary electrons in the hadronic scenario. To facilitate comparison with actual observations, we employ parameters that produce comparable X-ray and $\gamma$-ray spectra in both scenarios. Furthermore, to better approximate practical conditions, we consider the case in which a molecular cloud is present near the proton injection site (i.e., the spherical symmetry of the system does not hold). We investigate the resulting X-ray morphology and photon index variations in this environment and contrast them with the leptonic scenario.

\subsection{Spatial Distribution of Secondary electrons}

If protons and electrons are injected at the same position, the density distribution of propagated protons is supposed to exhibit a shallower gradient than that of primary electrons with the same diffusion coefficient, unless the energy loss of electrons is negligible. The spatial distribution of secondary electrons at injection basically follows that of parent protons if the density of the target for $pp$ collisions is homogeneous.  Subsequent diffusion of secondary electrons will further smooth their distribution. Therefore, we may expect a distinct distribution between primary electrons in the leptonic scenario and secondary electrons in the hadronic scenario.

To verify this point, we calculate the density distribution of secondary electrons at $t=2 \rm Myr$ for different magnetic field strengths with $B=10 \mu \rm G$, $ 50 \mu \rm G$ and $100\mu \rm G$. We assume the diffusion coefficient normalization $D_0$ to be $10^{26}~ \rm cm^2 s^{-1}$, which is significantly lower than the typical value in the ISM but consistent with the inferred value around some observed CR sources \citep{2017Sci...358..911A,2021NatAs...5..465A,2024SciBu..69..449L}. 
\begin{figure}[htb!]
\centering
    \gridline{\fig{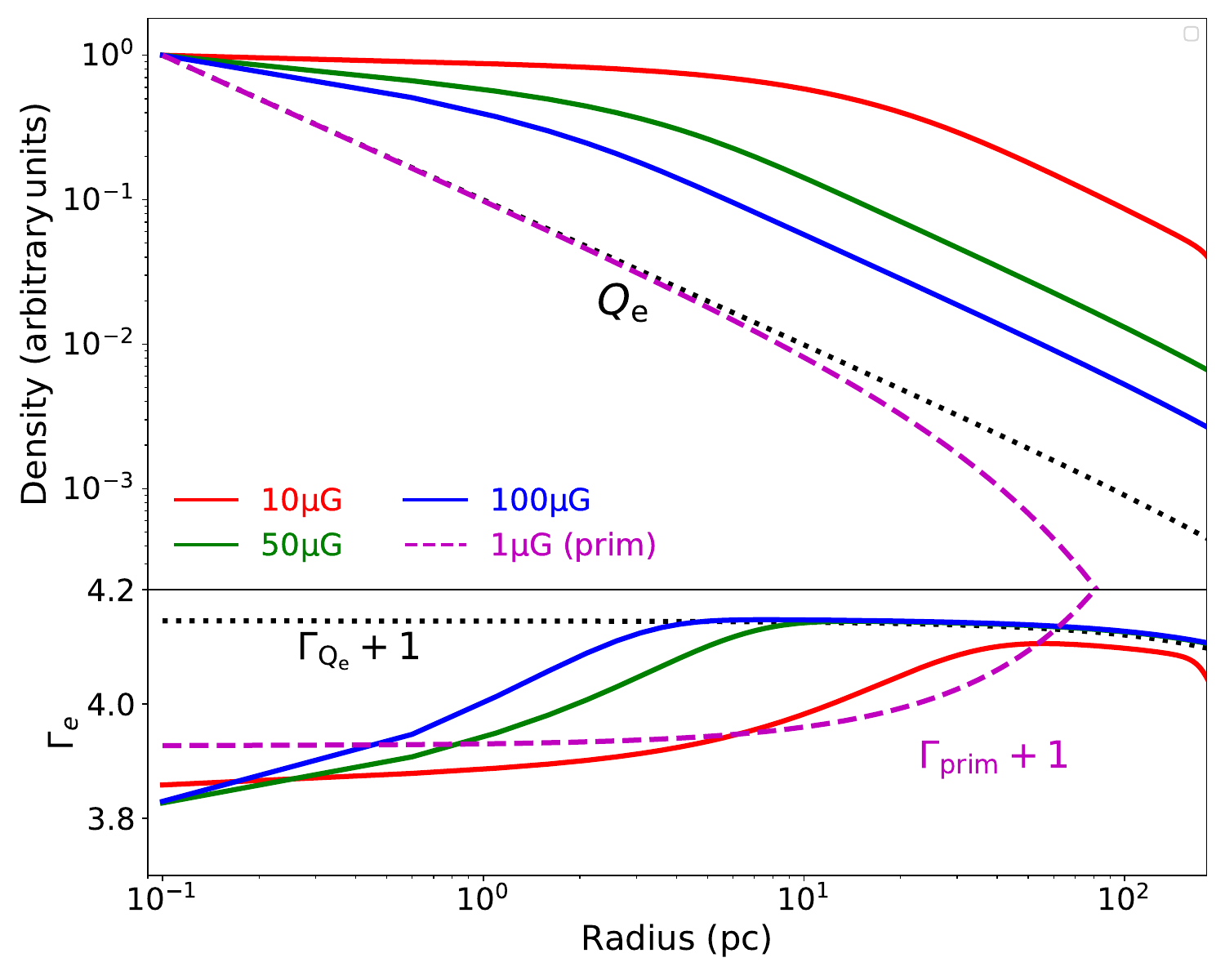}{0.48\textwidth}{}}
    \caption{Normalized electron density distribution (upper panel) and the power-law index profile (lower panel) of secondary electrons in $20-100$\,TeV.
    Different colors correspond to the magnetic field $B=10, 50, 100 ~ \mu$G. The purple dashed line corresponds to the normalized primary electron density in the leptonic scenario at $1 \, \rm \mu G$. The black dotted line corresponds to the normalized distribution of injection rate of secondary electrons $Q_{\rm e}$ at $t=2 \rm Myr$. Both the density and injection rate are normalized at $r= 0.1\,$pc.}
    \label{fig1}
\end{figure}

Figure~\ref{fig1} shows the predicted spatial evolution of density and power-law index of propagated electrons. The secondary electron density distribution exhibits a flat profile at small radii and steepens to $1/r$ at large radii. This transition occurs because at small radii the density is determined by a combination of electrons diffusing from the central high-density region and those injected locally, leading to a relatively flat profile. At larger radii, locally injected electrons dominate the density distribution, causing it to steepen. The distribution of injection rate $Q_e$ is shown in the same figure for reference. The injection of secondary electrons follows a $1/r$ distribution, reflecting the underlying proton distribution. We see that the stronger the magnetic field is, the closer the secondary electron density distribution follows that of $Q_e$, as enhanced synchrotron cooling restrains the diffusion of these high-energy electrons. For comparison, we also show the distribution of primary electrons assuming a weak magnetic field of $B=1\,\mu$G. We see that even with such a weak magnetic field, the density distribution of primary electrons becomes significantly steeper at large radii due to cooling.

Additionally, we analyze the power-law index of the electrons in 20-100 TeV in Figure \ref{fig1}. The spectrum softens with radius up to a certain distance before gradually hardening. This transition radius decreases with increasing magnetic field strength and is to some extent correlated with the radius where the electron density steepens. This behavior further supports our discussion on the distribution of electron density. At small radii, the spectral softening results from the diffusion of electrons outward from the central high-density region while simultaneously undergoing synchrotron cooling. At larger radii, the spectral hardening reflects the hardening of both injection and proton spectra. This occurs because high-energy protons diffuse farther within the system's age.

\subsection{Synchrotron radiation and its application}

\begin{figure*}[htb]
\centering
    \gridline{
    \fig{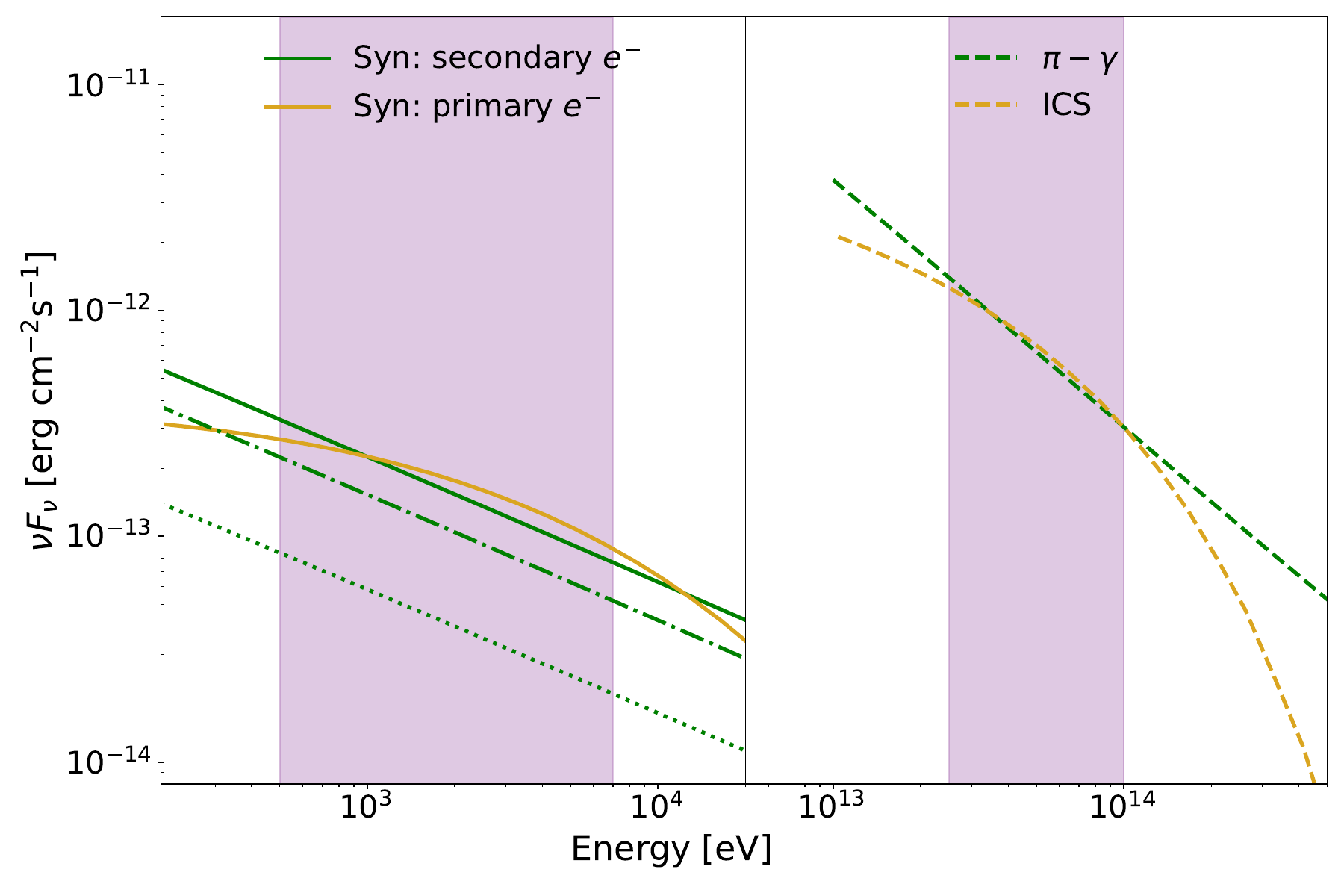}{0.50\textwidth}{}
    \fig{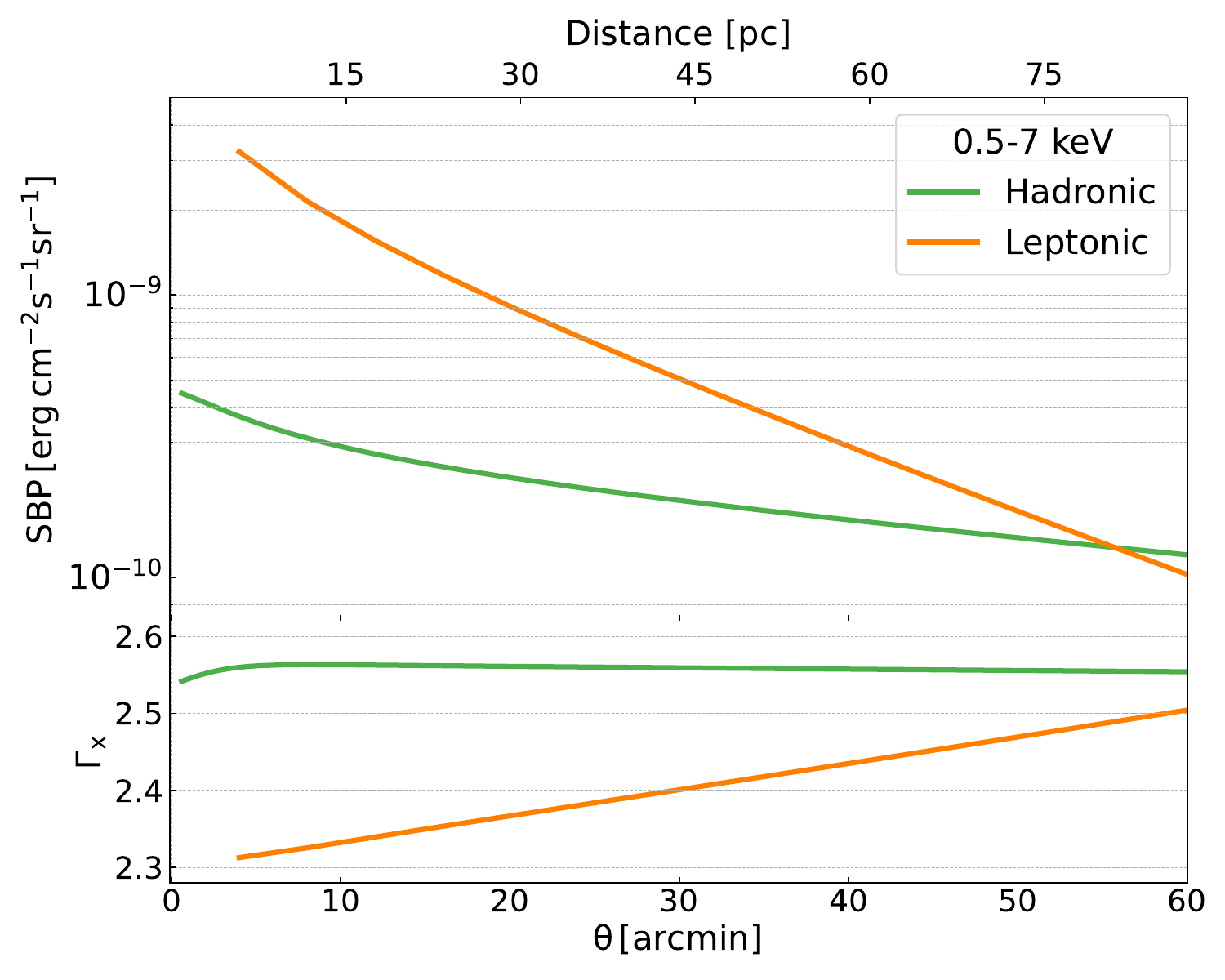}{0.44\textwidth}{}}
    \caption{Left: Energy spectrum for hadronic and leptonic scenario. In $\gamma$-ray band, both models satisfy $dN/dE_{\gamma} \propto E_{\gamma}^{-3}$ and $F_{25-100 \text{TeV}}= 1 \times 10^{-12} ~ \rm erg~cm^{-2}s^{-1}$. In the X-ray band, the yellow solid line refers to the synchrotron radiation from primary electrons with $B_{\rm lep}=1 \rm \mu G$. Green solid, dotted-solid and dotted lines refer to the synchrotron radiation from secondary electrons with $B_{\rm had}=100 \rm \mu G$, $50 \rm \mu G$ and $10 \rm \mu G$ respectively. Right: The predicted SBP and photon index profile in 0.5-7 keV with $B_{\rm had}=50 \rm \mu G$, $B_{\rm lep}=1 \rm \mu G$. The diffusion coefficient is assumed to $D_{\rm 0,had}$=$D_{\rm 0,lep}$=$10^{26} \rm cm^2/s$.}
    \label{fig2}
\end{figure*}

In this section, we present our results on the synchrotron radiation and its application to distinguish the radiation origin.

Previous works mainly focused on the energy spectra to distinguish between hadronic and leptonic origins. However, we argue that this approach may be insufficient in certain cases. Assuming $D_0=10^{26}~\rm cm^2 s^{-1}$ for both hadronic and leptonic scenarios, we first calculate the multi-wavelength spectrum, shown as the left panel of Figure \ref{fig2}. We assume the flux between 25-100 TeV follows $dN/dE_{\gamma} \propto E_{\gamma}^{-3}$ and $F_{25-100 \text{TeV}}= 1 \times 10^{-12} ~ \rm erg~cm^{-2}s^{-1}$. This requires the injection spectrum of $s_p=2.7$ for the hadronic scenario and $s_e=2.2$, $E_{cut}=200 ~\text{TeV}$ for the leptonic scenario. The required injection luminosity is $L_p (E_p > 100 \rm ~ TeV) \approx 10^{36} ~\rm erg/s$ in the hadronic scenario and $L_e(E_e > 100 \rm ~ TeV)\approx 10^{34} ~\rm erg/s$ in the leptonic scenario. For X-ray radiation in the leptonic scenario, we assume the magnetic field $B_{\text{lep}}=1 \mu \rm G$ to obtain a lower X-ray flux while keeping the flux due to ICS at the level of that from $pp$ collisions. For comparison, we present the hadronic scenario with $B_{\rm had} = 100 \mu \rm G$, $50 \mu \rm G$, and $10 \mu \rm G$, shown as solid, dash-dotted, and dotted lines, respectively. Overall, the hadronic and leptonic scenarios could predict similar fluxes both in $0.5-7$ keV and $25-100$ TeV with certain parameters, which suggests that energy spectra alone may not be sufficient to clearly distinguish between the two scenarios.

In this case, obvious differences in the energy-dependent morphology and photon index may provide valuable clues to the underlying emission mechanism. Figure~\ref{fig2} (right panel) shows the corresponding 0.5-7 keV SBP with $D_0 = 10^{26} ~ \rm cm^2 s^{-1}$ and $B_{\rm had}=50 \rm \mu G$, $B_{\rm lep}=1 \rm \mu G$. The hadronic scenario predicts a significantly flatter profile compared to the leptonic model, and a similar trend is observed in their photon indices. This difference arises from the cooling of primary electrons and is significant enough to distinguish between the two scenarios. Notably, this flatter profile extends far beyond the field of view of XMM-Newton and Chandra, indicating an extended X-ray background that might affect the derived flux in data analysis. In such cases, observations from more distant regions may be required to properly account for the X-ray background.

\begin{figure*}[htb]
\centering
    \gridline{\fig{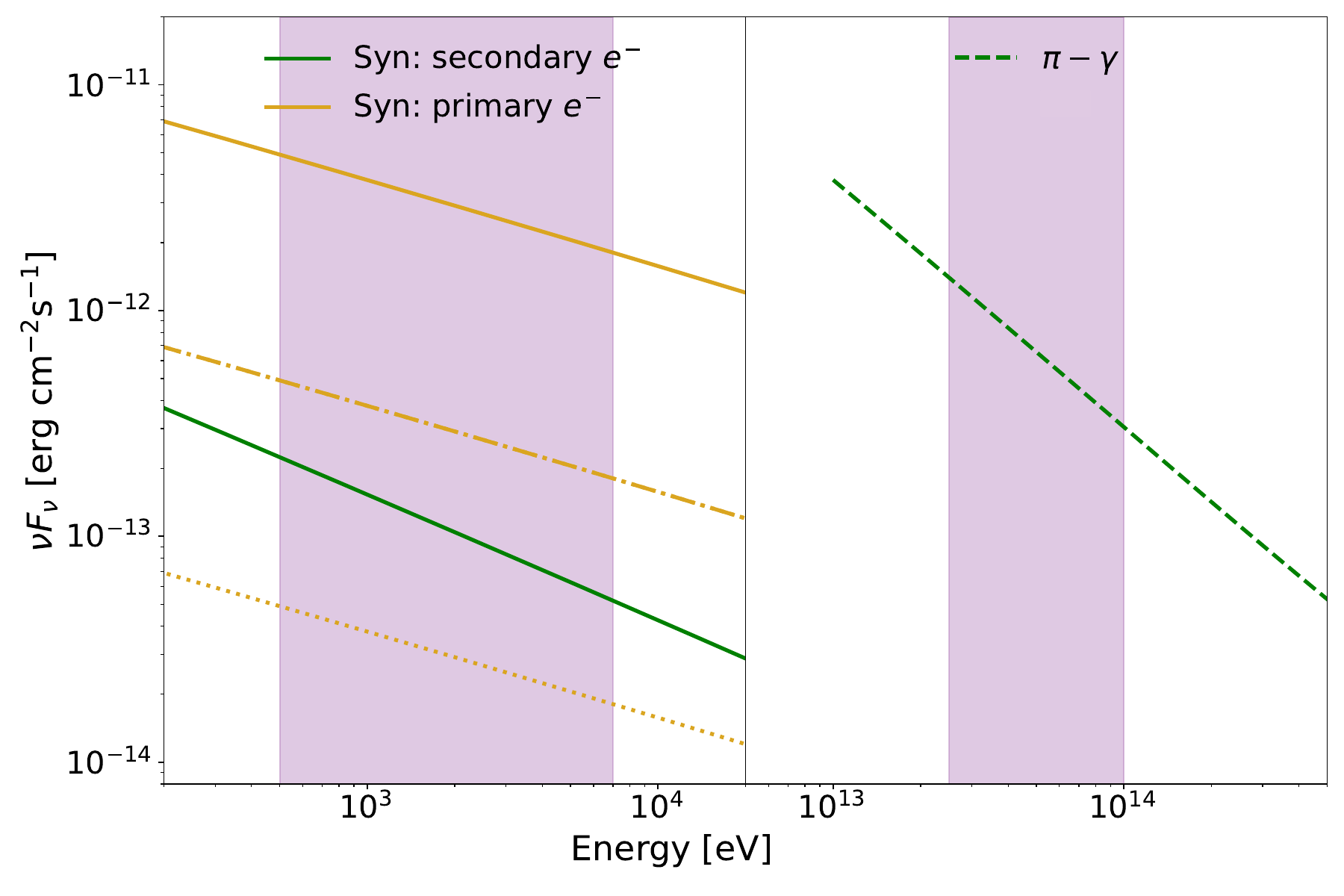}{0.50\textwidth}{}
    \fig{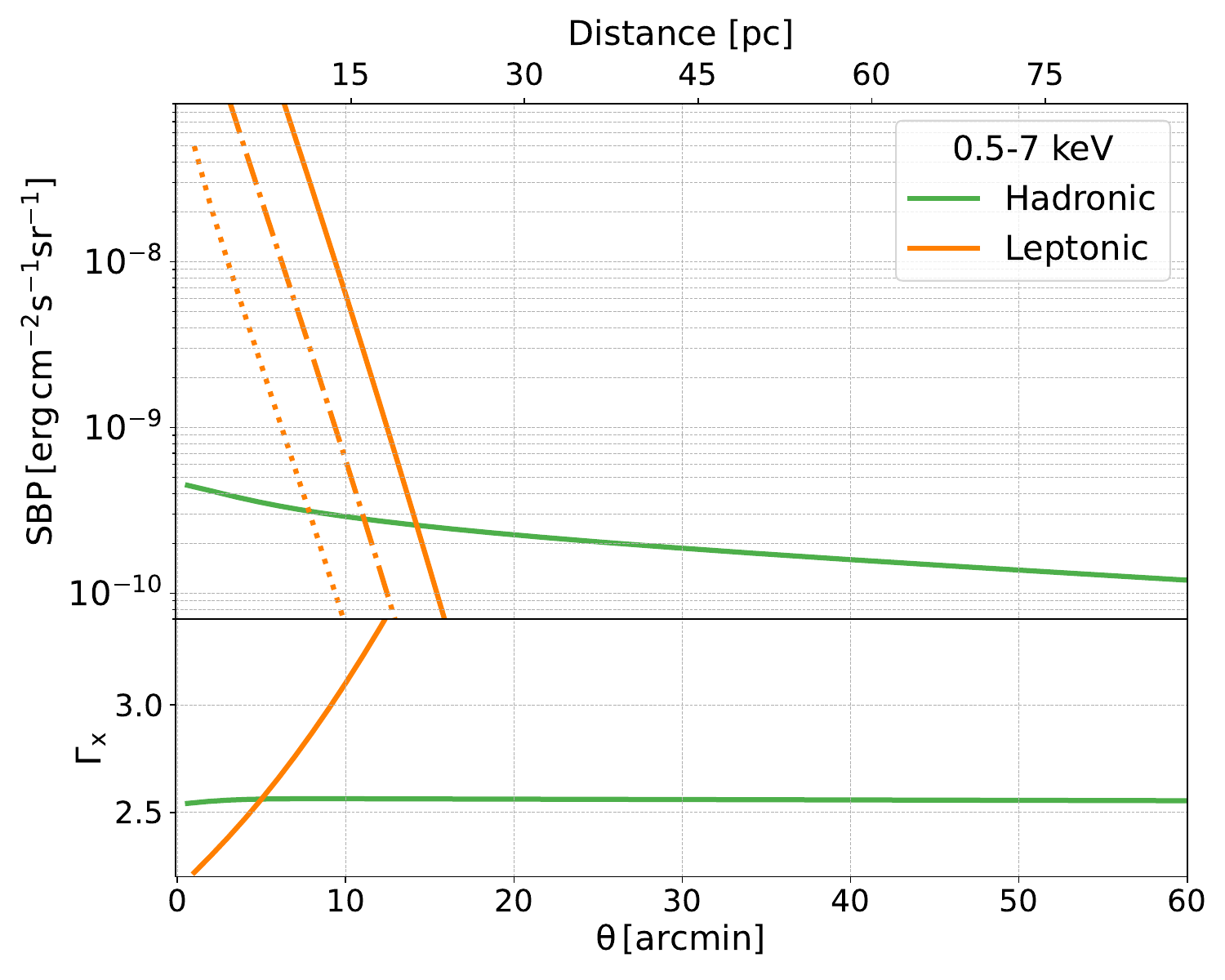}{0.44\textwidth}{}
    }
    
    \caption{The predicted energy spectrum, SBP and photon index profile in the hybrid case with $B_{\rm hybrid}=50 \rm \mu G$ and $D_{\rm 0,had}$=$D_{\rm 0,lep}$=$10^{26} \rm cm^2/s$. The solid, dash-dot and dotted lines refer to $K_{ep}=10^{-2}, 10^{-3}$ and $10^{-4}$.}
    \label{fig:hybrid}
\end{figure*}

Additionally, we would like to note that the X-ray intensity profile and the photon index profile generated by secondary electrons are calculated in an idealized scenario, i.e., no contamination from primary electrons and thermal emissions. Indeed, even for a hadronic $\gamma$-ray source, electrons must be accelerated along with protons, and a modest amount of primary electrons could dominate the X-ray flux of the source. Hence, we conduct additional tests on this hybrid case, where both accelerated protons and electrons exist, and the TeV emission is dominated by the hadronic process. Assuming an identical injection spectrum $s_e=s_p=2.7$, we define $K_{ep}$ as the ratio of the differential number density of injected primary electrons to that of protons. Consequently, we have $Q_e = K_{ep} Q_p$. Typically, $K_{ep}$ can vary from $10^{-4}$ to $10^{-2}$ inferred from the observed CR flux, individual young supernova remnant and kinetic simulation \citep{2006A&A...451..981B,2012A&A...538A..81M,2013ApJ...767...20L,2015PhRvL.114h5003P}. We choose $K_{ep}=[10^{-4},10^{-3},10^{-2}]$ and $B_{\rm hybrid}=50\mu \rm G$ to compute the corresponding emission. Figure \ref{fig:hybrid} shows the energy spectrum and SBP, the X-ray flux is dominated by primary electrons for $K_{ep} >4\times 10^{-4}$. For the spatial distribution, we observe a transition from steep to flat brightness profile for all the values of $K_{ep}$. The transition radius varies from 8 to 15 arcmin. This means an extended tail with an almost constant photon index profile appeared in the outer region, which falls within the FoV of some wide-field X-ray detectors (e.g., Einstein Probe and eROSITA). Hence, the X-ray profiles could still reveal the secondary electrons and the hadronic component.

\subsection{Radiation distribution considering the presence of molecular clouds}

In the previous sections, we performed calculations and discussions under the assumption of a uniform density medium, representing the contribution from relatively dense regions of the ISM. However, in the vicinity of CR sources, denser molecular clouds or clumps may exist, leading to a localized hotspot of emission. The corresponding $\gamma$-ray and multi-wavelength emissions have previously been investigated theoretically from the perspective of their spectra \citep{1996A&A...309..917A,2005A&A...432..609B,2009MNRAS.396.1629G}. In this section, we introduce an offset molecular cloud component and evaluate its impact on the spatial distribution of synchrotron radiation. The density profile of the molecular cloud is assumed to follow the Plummer function \citep{1911MNRAS..71..460P, 2014A&A...565A..24F}: 

\begin{equation}
\resizebox{0.5\columnwidth}{!}{$
\begin{aligned}
n_{\rm MC}(r')=\frac{n_{0}}{[1+(r'/r_{c})^2]^{\frac{\alpha}{2}}}
\end{aligned}
$}
\label{eq:mc}
\end{equation}
where $n_0$ is the density at $r'=0$, $r_c$ is the radius of the core region, and $\alpha$ is the Plummer index. Note that $r'$ represents the distance to the center of the molecular cloud, while the center of the molecular cloud is assumed to be offset by 30\,pc from the proton injection center, i.e., $r'^2=(z-30\,\text{pc})^2+r^2$. The Plummer function describes a density distribution characterized by a flat core and a steep decline in the outer region, making it well-suited for modeling molecular clouds. In this paper, we assume $\alpha=4$ and $r_c=10 \rm ~pc$, so that the mass is concentrated in the central $10 ~\rm pc$ region. $n_0$ is assumed to be $1000 \rm ~ cm^{-3}$, corresponding to the mass of the core region $M_{\rm core}=9 \times 10^{4}~ M_{\rm sun}   $\footnote{The average number of collisions experienced by a CR proton while traversing the MC is $N_C = c \sigma_{pp} r_c^2 n_0 /3 D(E_p) \textless \textless 1$ for $E_p \textless  1 ~ \rm PeV$, which enables us to ignore the energy loss of protons. For a detailed discussion, refer to \citet{ge2025impact}.} and column density $N_{\rm H_2}=2.4\times 10^{22} ~\rm cm^{-2}$. The high density in the core region will lead to strong absorption of soft X-ray photons. A more detailed calculation would involve considering the absorption cross-section for photoionization and other processes \citep{2000ApJ...542..914W}. However, this is not the focus of the current work. Therefore, we neglect absorption and focus on the X-ray radiation in the $2-7$\,keV range, as higher energy radiation is less affected by absorption.

We adopt the same injected proton spectrum and luminosity as in Section 3.2. In this case, the injected protons diffuse and interact with both ISM and the molecular cloud, producing two distinct populations of secondary electrons: One arising from collisions between CR protons and ISM, the other from collisions between CR protons and MC. The total injection and thus the final distribution of secondary electrons can be expressed as: $N_{e, \rm All} = N_{e, \rm ISM}+ N_{e, \rm MC}$. Here, $N_{e, \rm ISM}$ has been derived in the previous section, and $N_{e, \rm MC}$ remains to be determined. Due to the offset of the molecular cloud, $N_{e, \rm MC}$ is no longer spherically symmetric with respect to the particle injection center, but instead exhibits symmetry along the axis connecting the proton injection center and the center of the molecular cloud. Therefore, we calculate the injection and diffusion of electrons in cylindrical coordinates and obtain $N_e(r,z,E_{\rm e})$. The origin ($r=0, z=0$) is located at where protons are injected, while the center of the molecular cloud is located at $r=0, z=30$ \,pc. Finally, we obtain the synchrotron radiation flux and integrate the flux along the line of sight to get the intensity $I_{\rm MC}$ and the total intensity $I_{\rm All}=I_{\rm MC}+I_{\rm ISM}$. For comparison, we also calculate the 2D intensity distribution of $\gamma$-ray emission from $pp$ collisions between protons and the target medium including ISM and molecular cloud. In the calculation, we fix $n_{\rm ISM}=10 ~ \rm cm^{-3}$, and the interstellar magnetic field strength is fixed at $B_{\rm ISM}=3 \,\rm \mu G$. To model the magnetic field strength within the MC, we assume the relationship $B \propto n^{1/2}$ \citep{2015MNRAS.451.4384T}, although earlier work by \citet{2010ApJ...725..466C} proposed $B \propto n^{2/3}$, and use a distribution represented as:
\begin{equation}
\resizebox{0.6\columnwidth}{!}{$
\begin{aligned}
B_{\rm MC}(r')=50 ~\left(\frac{n_{\rm MC}}{10^3 ~ \rm cm^{-3}} \right)^{\frac{1}{2}} \mu \rm G,
\end{aligned}
$}
\label{eq:Bvar}
\end{equation}
The final magnetic field is obtained as $B(r,z)=\max(B_{\rm MC}(r,z), B_{\rm ISM})$.


\begin{figure*}[htb!]
\centering
    \gridline{
       \fig{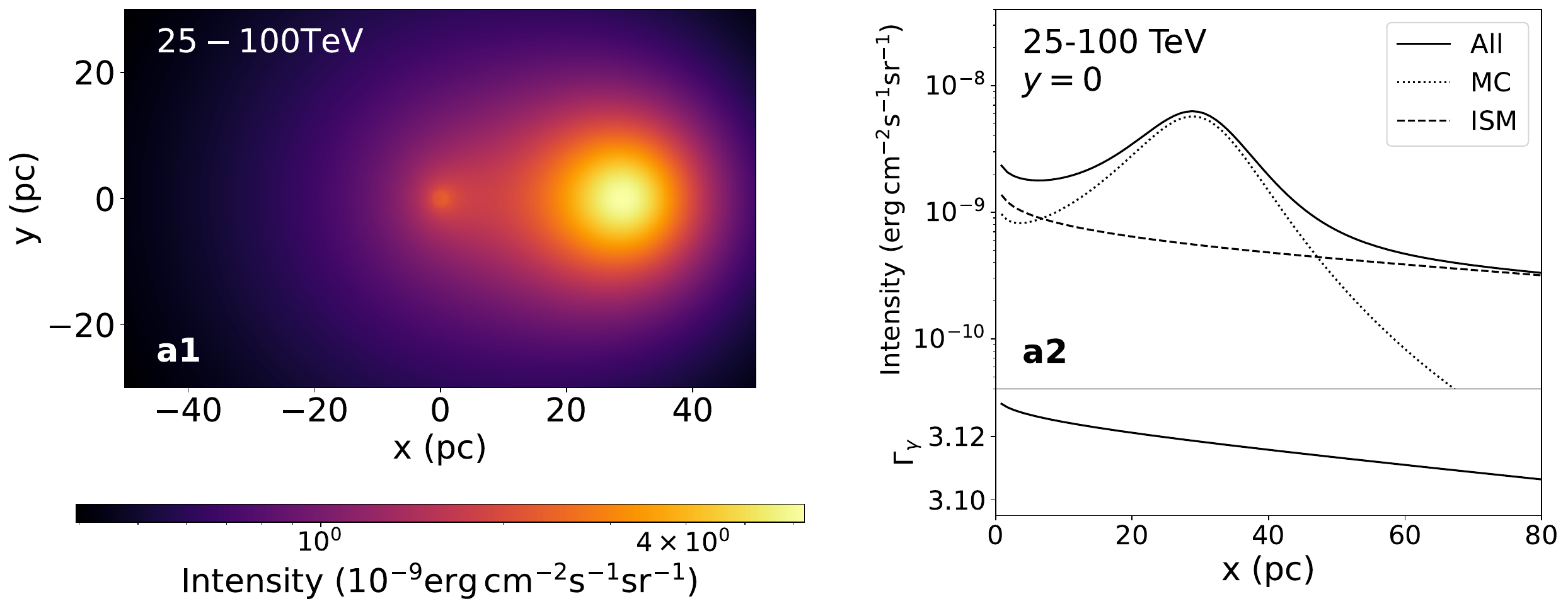}{0.8\textwidth}{}
    }\vspace{-0.5cm}
   \gridline{
       \fig{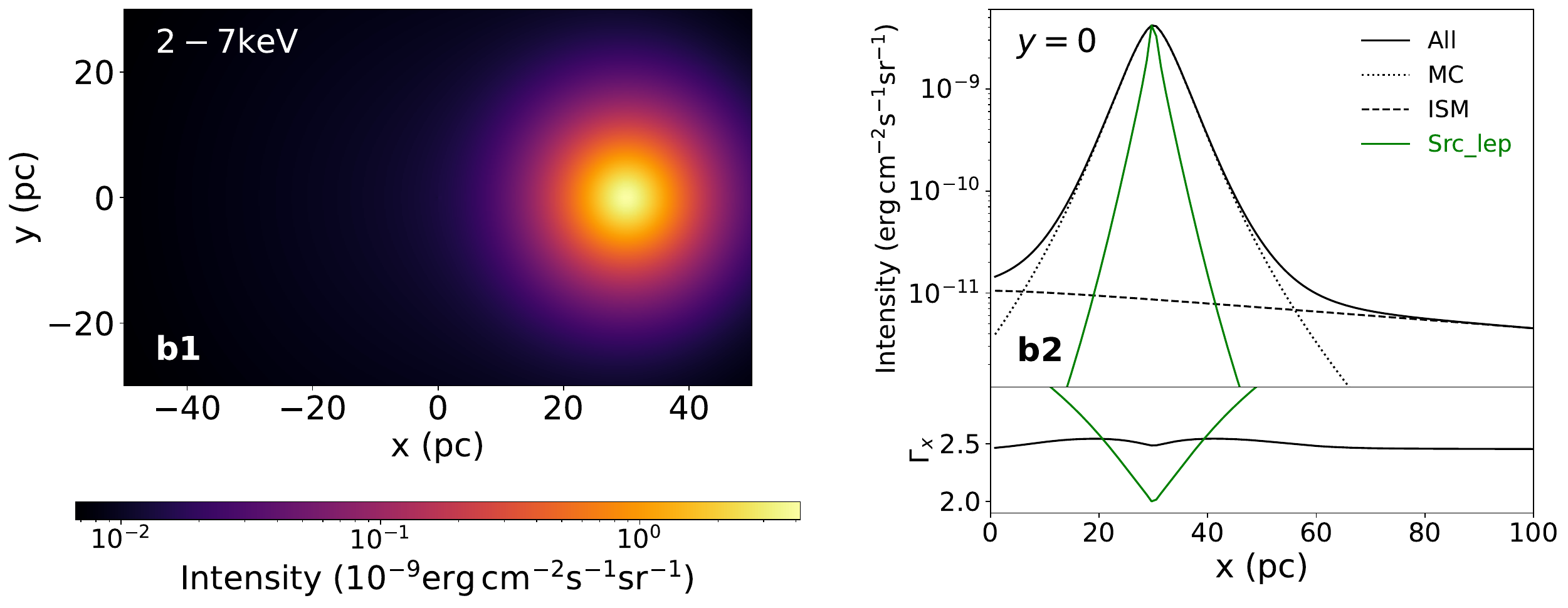}{0.8\textwidth}{}
    }

\caption{Left: the morphology of emission in hadronic scenario. Right: the corresponding SBPs and photon index profiles along the x-axis in hadronic (black lines) and leptonic (green lines) scenarios. From top to bottom, (a) $25-100$\,TeV $\gamma$-ray emission (b) $2-7$\,keV X-ray emission.}

\label{fig:map_sbp}
\end{figure*}

Figure \ref{fig:map_sbp} presents our calculated results. The left column shows the 2D morphology of the radiation, while the right column displays the 1D SBPs and the photon index profiles along the line connecting the proton injection center and the molecular cloud center. The top row of Figure \ref{fig:map_sbp} displays the result in the $\gamma$-ray ($25-100$\,TeV) band, and the bottom row showcases the result in the X-ray ($2-7$\,keV) band. From the top row (panels a1 and a2), we see two peaks in the $\gamma$-ray band both from the 2D map and the 1D SBP: one relatively bright compact source centered at the MC center, one relatively faint extended source centered at the proton injection location. The former one is due to the high density of the target gas inside the MC, while the latter is due to the high density of relativistic particles around the injection center. In the X-ray band, however, we did not see an apparent source around the injection center, because of the lower magnetic field in ISM (see Figure~\ref{fig:map_sbp}b).  It may be worth further discussing the difference in the predicted $\gamma$-ray morphology and the X-ray morphology. The X-ray source centered at the MC in both the X-ray 2D morphology and 1D profile appears more compact compared to that in the $\gamma$-ray map. This phenomenon can be attributed to the decline of the magnetic field strength at the outer part of the MC.

In the bottom row of Figure 4, we observe an X-ray source centered at the MC, produced by the synchrotron radiation of secondary electrons. One may also wonder that if in reality an X-ray source is observed in association with an MC, whether we can identify its hadronic origin because the X-ray source could also be produced by an electron accelerator inside the MC or even a background/foreground electron accelerator outside the MC. We therefore test the leptonic scenario by injecting electrons in molecular cloud at $r=0, z=30$ pc (Src$_{\rm lep}$). We adopt the same injection spectrum as in Section 3.2 ($s_e=2.2, E_{\rm cut}= 200$\,TeV). The setup of the magnetic field remains the same as in the hadronic scenario. The electron injection luminosity is tuned to make the peak X-ray intensity generated in the leptonic scenario the same as that in the hadronic scenario, so that we can compare the intensity profile more straightforwardly. The value is $L_{e,\rm MC}(E_e>100 ~\rm TeV)= 8.7\times 10^{30} ~$erg/s. The green line in Figure~\ref{fig:map_sbp} (b2) shows the corresponding 1D SBPs and photon index profiles along the x-axis in 2-7 keV in the leptonic scenario. We see the intensity profile has a steeper decline in the leptonic scenario than the hadronic scenario due to the rapid cooling of escaping electrons. More notably, the photon index profile presents a rapid softening in the leptonic scenario. Even if the electron accelerator is located outside the MC (at the foreground/background but coincident with the light of sight of the MC), where the magnetic field could be much weaker than $B_{\rm MC}$, the softening in the X-ray photon index profile can be still quite pronounced compared to the hadronic scenario as discussed in Section 3.2 (see also Figure~\ref{fig2}). This can serve as a distinguishing feature between the hadronic scenario from the leptonic scenario.

\section{Discussion and Conclusion} \label{sec:dis}
In this paper, we consider a source with a distance of $d=5 \rm kpc$ and a suppressed diffusion coefficient. This allows the majority of the flux to be concentrated within a region smaller than $1^\circ$, which is not much bigger than the field of view of some X-ray instrument (such as Einstein Probe and eROSITA). This would be advantageous for X-ray detection. In the case that sources are closer to us or the diffusion coefficient is larger, the radiation region may be more extended. The size of the source may exceed the field of view of instruments, causing them to capture only a small portion of the flux without multiple pointings. Furthermore, the flat SBP could complicate X-ray analyses, because it would be more difficult to distinguish the emission of the source itself from the background.

This study assumes that CRs are injected from a point-like source. In reality, CRs may be injected from the surface of, for example, a shell-type or a spherical geometry. The resulting particle distribution would be slightly different than what we obtain here, but the main differences in the SBPs and photon index profiles between the hadronic and leptonic scenarios should remain more or less the same. This is because the differences are mainly due to the slower cooling of protons compared to electrons and the diffusion of secondary electrons.

Additionally, we only consider the isotropic diffusion of CRs within a region of homogeneous diffusion coefficient. Additional transport processes, such as anisotropic diffusion or the self-triggered confinement of CRs, would further complicate the expected morphology of the radiation. For instance, in the case of anisotropic diffusion, the morphology also depends on the local turbulence level and the angle between the line of sight and the local mean magnetic field direction \citep{2019PhRvL.123v1103L}. In the context of pulsar halos, the constraints provided by observations on anisotropic models indicate a rapidly declining profile, suggesting a suppressed diffusion coefficient perpendicular to the line of sight. On the other hand, the self-excited turbulence by CRs, which could result in a spatially dependent diffusion coefficient, might further complicate the radiation morphology. Near the injection location, the higher CR density amplifies turbulence more efficiently, leading to a reduced diffusion coefficient and a more concentrated morphology. Nevertheless, our main conclusion holds qualitatively. 

Finally, we note that the secondary and primary electron models require distinct environmental conditions, particularly in terms of magnetic field strength and medium density. Our calculations show that, to produce observable X-ray emission, the hadronic model, involving secondary electrons from $pp$ collisions, necessitates a dense target medium and a magnetic field strength at least at $B \sim 10 \, \mu\text{G}$ level. In contrast, the leptonic model (e.g., in pulsar halos) can operate in lower-density environments with weaker magnetic fields ($B \sim 1-10 \, \mu\text{G}$). However, in regions with strong magnetic fields ($B > 100 \, \mu\text{G}$), the synchrotron emission from a small population of primary electrons could dominate the SED in the X-ray band, potentially overshadowing the hadronic contribution. If magnetic field strengths or medium densities can be constrained a priori through observations, these parameters provide a complementary diagnostic tool to differentiate between the two scenarios.

To summarize, we investigated the spatial morphology of X-ray synchrotron radiation produced by secondary electrons generated in $pp$ collisions, and compared the results with those produced by primary electrons. In our calculation, protons are injected from a point source and diffuse isotropically, interacting with the surrounding medium. Our main findings include:

(i) We calculated the density profile of secondary electrons, which exhibits a flat profile at small radii and steepens to $1/r$ at large radii. We attribute this behavior to the mixing of diffusing electrons from the high-density central region with locally injected electrons. electrons power-law index profile supports this explanation.

(ii) We calculated the multiwavelength emission in both the hadronic and leptonic scenarios. Given only the X-ray and $\gamma$-ray spectrum measurements, we found that it could still be challenging to determine the radiation mechanism of $\gamma$-ray photons purely based on the energy spectrum. Instead, intensity profiles and photon index profiles may provide additional information and help differentiate the radiation origin. Additionally, a flat X-ray intensity profile as predicted in the hadronic scenario may confuse the radiation of the source itself and the background radiation. Therefore, we need to be cautious in the background selection of the X-ray analysis for UHE $\gamma$-ray sources.

(iii) We considered the presence of an offset molecular cloud, observing a pronounced correlation between the X-ray radiation and the molecular cloud. The profile steepens near the molecular cloud and flattens in the outer region. This steep profile could be similar to that predicted in the leptonic scenario, but the photon index exhibits no substantial cooling effect and can serve as a diagnostic tool. What's more, $\gamma$-ray and X-ray emissions may exhibit completely different spatial morphologies, which could potentially provide a theoretical explanation for certain sources in multiwavelength observations.


\section*{Acknowledgements}
This work is supported by the National Natural Science Foundation of China under grants No.~12393852 and 12333006.

\vspace{5mm}

\bibliography{sample631}{}
\bibliographystyle{aasjournal}



\end{document}